\documentstyle[preprint,aps,psfig]{revtex}
\begin{document}
\draft
\title{Decoherence in ion traps due to laser intensity and phase fluctuations}
\author{S.~Schneider and G.~J.~Milburn}
\address{Department of Physics,
The University of Queensland,
QLD 4072 Australia.}
\maketitle

\begin{abstract}
We consider one source of decoherence for a single 
trapped ion due to intensity and phase fluctuations in the 
exciting laser pulses.  For simplicity we assume that the stochastic 
processes involved are white noise processes, which enables us to give a 
simple master equation description of this source of decoherence. 
This master equation is averaged
over the noise, and is sufficient to describe the results 
of experiments that probe the oscillations in the electronic 
populations as energy is exchanged between the internal and
electronic motion. Our results are in good qualitative agreement 
with recent experiments and predict that the decoherence rate will 
depend on vibrational quantum number in different ways
depending on which vibrational excitation sideband is used. 
\end{abstract}

\newpage
\section{Introduction}
Recent advances in laser cooling  now enable a 
single trapped ion to be 
prepared in a chosen quantum state of the center-of-mass vibrational 
motion\cite{Wineland1997}. In some cases, the state is highly nonclassical, 
such as the 
recent preparation of a  superposition of two oscillator coherent 
states\cite{Monroe1996}. Quantum dynamical features, such as collapse and 
revival oscillations, have also been observed\cite{Wineland1996}.
The key innovation in such experiments is the ability to 
tailor the effective potential experienced by the ion by  coupling the 
center-of-mass motion to the electronic states by external 
laser pulses. If more than a single ion is trapped, individual ions may 
be addressed by different laser pulses leading to an entanglement of the 
collective vibrational motion of the ions and their electronic states. 
This is the principle behind the suggestion of Cirac and Zoller
\cite{Cirac1995} for 
implementing quantum computation in ion traps. So far however only  a 
single controlled-NOT gate, a key component in a quantum computer, has 
been implemented experimentally \cite{Monroe1995}.  

Despite these heroic 
experimental achievements, the quantum motion of a single trapped ion is 
obviously limited by sources of decoherence. Decoherence arises from 
random and unknown perturbations of the Hamiltonian. If these 
perturbations cannot be followed exactly, experiments must average over 
them. This leads to an effective irreversible evolution of the trapped 
ion and a suppression of coherent quantum features through the decay of 
off-diagonal matrix elements of the density operator in some basis. 
Complementary to the decay of off-diagonal matrix elements, noise is added 
to conjugate variables. This can appear as a heating of the ion if noise is
added to the 
momentum variable.    

In this paper we consider 
one source of decoherence for a single 
trapped ion due to intensity and phase fluctuations in the 
exciting laser pulses.
For simplicity we assume that the stochastic 
processes involved are white noise processes, which enables us to give a 
simple master equation description of this source of decoherence. 
Section II contains a general overview of the kind of system we 
investigate. In the first subsection we concentrate on intensity 
fluctuation in the laser, whereas the second subsection is devoted 
to phase fluctuations. We conclude with a discussion on the 
experimental relevance of our results.

\section{Laser fluctuations.}

A single two-level ion, with mass $m$,  tightly bound in a 
harmonic trap, and laser cooled to the 
Lamb-Dicke limit, can be prepared in a variety of states by carefully 
controlling the effective detuning of external laser fields which couple 
the vibrational motion and the internal electronic states. 
For simplicity we will assume the ion is constrained to
move in a single dimension at harmonic frequency $\nu$. A reference 
frequency is provided  by the atomic transition frequency 
$\omega_A$. If the effective laser  frequency $\omega_L$ is tuned
below or above this frequency by multiples of the harmonic trap 
frequency $\nu$, a variety of effective potentials may be obtained. 
In the NIST experiments \cite{Wineland1997}, two laser fields
are used to excite two-photon stimulated Raman transitions. 
However in this paper we will consider the simplest case
of a single classical laser, with wave vector $k_L$ and frequency $\omega_L$, 
where the field is propagating in
the same direction in which the ion is constrained to vibrate.

In the interaction picture with the dipole and rotating wave 
approximation the interaction Hamiltonian is\cite{Wineland1997} 
\begin{equation}
H_I=\hbar\Omega(t)(\sigma_++\sigma_-)\left (e^{i\eta(a^{-i\nu t}+a^\dagger
e^{i\nu t})-i\delta
t+i\phi(t)}\right )+h.c \,\, ,
\label{hamiltonian}
\end{equation}
where $\Omega(t)$ is the effective Rabi frequency for this 
transition, written as a function of time to account for 
fluctuations resulting from laser intensity fluctuations, $\phi(t)$ 
represents fluctuations in the laser phase, 
$\eta=k_L(\hbar/2m\nu)^{1/2}$ is the Lamb-Dicke parameter,
$\delta=\omega_A-\omega_L$ is the detuning between the laser 
and the electronic states, and $\sigma_\pm$ are the usual 
two-level atom transition operators. In order to excite particular
transitions, $|g,n\rangle \leftrightarrow |e,n^\prime\rangle$, 
of the coupled electronic/vibrational spectrum we choose the detuning 
\begin{equation}
\delta=\nu(n-n^\prime) \,\, ,
\end{equation}
where $n,n^\prime$ are integers. In this paper we will assume that 
the amplitude of the ions motion in the direction of the 
laser field is much less than a wavelength. In this limit we can
expand the interaction Hamiltonian to lowest order in $\eta$
\cite{Wineland1997}. Furthermore, to illustrate the effect of 
laser fluctuations it will suffice to consider four cases: 
(i) carrier excitation, $n=n^\prime$, 
(ii) first red sideband excitation $n^\prime=n-1$,
(iii) first blue sideband, $n^\prime=n+1$,
(iv) second red sideband, $n^\prime = n+2$. 
The interaction Hamiltonians for these four cases are:
\begin{eqnarray}
\label{randomham1}
H_I(t) & = & \hbar\Omega(t)(1+\eta^2a^\dagger a)\left
[\sigma_+e^{i\phi(t)}+h.c\right ]\ \ \ \mbox{carrier} \,\, ,\\
H_I(t) & = & \hbar\Omega(t)\eta\left [\sigma_+ a e^{i\phi(t)}
+h.c\right ]\ \ \ \mbox{red sideband} \,\, , \\
H_I(t) & = & \hbar\Omega(t)\eta\left [\sigma_+ a^\dagger e^{i\phi(t)}
+h.c\right]\ \ \ \mbox{blue sideband}\,\, , \\
H_I(t) & = & \hbar\Omega(t)\eta\left [ \sigma_+ a^2 e^{i\phi(t)}
+h.c\right]\ \ \ \mbox{second red sideband}\,\, . 
\label{randomham}
\end{eqnarray}
The red sideband Hamiltonian corresponds to the familiar
Jaynes-Cummings Hamiltonian \cite{JC} of quantum optics. 

We will specify the noise by defining a stochastic process for 
$\Omega(t)$ and $\phi(t)$. In the
case of laser amplitude fluctuations we take
\begin{equation}
\Omega(t)dt=\Omega_0(dt+\sqrt{\Gamma}dW(t)) \,\, , 
\label{amplitude}
\end{equation}
where $dW(t)$ is the increment of a real Wiener 
process\cite{Gardiner1985}, and $\Omega_0$ is the
non-fluctuating component of the Rabi frequency. The parameter 
$\Gamma$ scales the noise. The interpretation of
$\Gamma$ is given by integrating Eq.(\ref{amplitude}) to obtain the pulse 
area, $A(T)$, which is also a stochastic variable,
\begin{equation}
A(T) =\Omega_0T+\Delta A(T) \,\, ,
\end{equation}
where $T$ is the pulse duration. The last term in this 
equation is a Gaussian random variable with mean zero and variance 
$E(\Delta A(T)^2)=\Omega_0^2\Gamma T$. If we then consider 
the ratio of the r.m.s. fluctuations in the pulse area
to the deterministic pulse area we find
\begin{equation}
\frac{E(\Delta A(T)^2)^{1/2}}{\Omega_0 T}=\sqrt{\frac{\Gamma}{T}} \,\, .
\end{equation}
For phase fluctuations we take a simple diffusion,
\begin{equation}
\phi(t)=\sqrt{\gamma}W(t) \,\, ,
\end{equation}
where $W(t)$ is the Wiener process. 

\subsection{Intensity fluctuations}

The Hamiltonians in Eqs.~(\ref{randomham}) are stochastic. We first 
consider the effect of laser intensity fluctuations and ignore 
phase fluctuations. We thus set $\phi(t)=0$ (constant).  To
obtain the corresponding Schr\"{o}dinger equation requires some care, 
as the white noise process is quite singular. We can however define 
a stochastic Schr\"{o}dinger equation in the  Ito
formalism\cite{dyrting1996}, or more appropriately a stochastic 
Liouville-von Neumann equation,
\begin{equation}
d\rho(t)=-i[G,\rho]dt-i\sqrt{\Gamma}[G,\rho]dW(t)
-\frac{\Gamma}{2}[G,[G,\rho]]dt\,\, , 
\end{equation}
where $G$ takes one of the four forms,
\begin{eqnarray}
\label{ggg1}
G & = & \Omega_0\sigma_x(1+\eta^2a^\dagger a)
\ \ \ \mbox{carrier} \,\, , \\
G & = & \eta\Omega_0(a\sigma_+ + a^\dagger \sigma_-)
\ \ \ \mbox{red sideband} \,\, , \\
G & = & \eta\Omega_0(a^\dagger\sigma_++a\sigma_-)
\ \ \ \mbox{blue sideband} \,\, , \\
G & = & \eta\Omega_0(a^2 \sigma_+ + (a^\dagger)^2 \sigma_-)
\ \ \ \mbox{second red sideband} \,\, .
\label{ggg}
\end{eqnarray}
This equation gives the evolution of the system density operator 
conditioned on a particular noise history.  In an experiment involving a 
number of pulses, with data from each pulse combined, the
noise is effectively averaged and we obtain the following master equation
describing the system
\begin{equation}
d\rho(t)=-i[G,\rho]dt-\frac{\Gamma}{2}[G,[G,\rho]]dt \,\, . 
\label{mastereq}
\end{equation}
This equation has a similar form to that considered in reference
\cite{milburn1991} for a model of intrinsic decoherence. Indeed, 
that model of intrinsic decoherence has been explicitly solved for
the case of the Jaynes-Cummings Hamiltonian (the red side-band case) by
Moya-Cessa \em et al.\em\cite{Moya} (see also \cite{Kuang}).

The last term in Eq.(\ref{mastereq}) is responsible for decoherence and,
complementary to decoherence, it leads to diffusive growth in 
observables that do not commute with $G$, which is
proportional to the interaction Hamiltonian of the system.  
In each case, the eigenstates of $G$
are discrete and labeled by an index $n$ corresponding to a 
vibrational quantum number, and a sign
index $\pm$ arising from the two dimensional Hilbert space 
of the electronic motion. We will designate these states as
$\{|e_n^\pm\rangle\}$. Thus
\begin{equation}
\frac{\partial}{\partial t}\langle e_n^\pm|\rho(t)|e_m^\pm\rangle=
\left [-i(e_n^\pm-e_m^\pm)-\frac{\Gamma}{2}(e_n^\pm-e_m^\pm)^2\right ]\langle
e_n^\pm|\rho(t)|e_m^\pm\rangle \,\, . 
\label{energyme}
\end{equation}
In this form the decay of off-diagonal coherence is quite explicit. 
Notice however that decoherence takes place in a joint basis of 
both the electronic and vibrational Hilbert spaces. This is in
contrast to many similar proposals for decoherence, such as Brownian motion,
which only involve a single Hilbert space. Furthermore, the form of 
the decoherence ensures that total energy is
conserved even in the  presence of noise, as expected for a stochastic
Hamiltonian. 

The eigenstates and eigenvalues for each of the operators in 
Eqs.(11-13) are as follows.
For the case of carrier excitation:
\begin{eqnarray}
|e_n^\pm\rangle & = & \frac{1}{\sqrt{2}}\left (|g,n\rangle\pm|e,n\rangle\right
)\\
e_n^\pm & = & \pm\frac{\hbar\Omega_0}{2}(1+n\eta^2) \,\, .
\end{eqnarray}
For the case of red side-band excitation (Jaynes-Cummings):
\begin{eqnarray}
|e_n^\pm\rangle & = &  \frac{1}{\sqrt{2}}\left
(|g,n+1\rangle\pm|e,n\rangle\right )\\
e_n^\pm & = & \pm\eta\Omega_0\sqrt{n+1} \,\, . 
\end{eqnarray}
For the blue side-band case:
\begin{eqnarray}
|e_n^\pm\rangle & = &  \frac{1}{\sqrt{2}}\left
(|g,n-1\rangle\pm|e,n\rangle\right )\\
e_n^\pm & = & \pm\eta\Omega_0\sqrt{n} \,\, . 
\end{eqnarray}
For the second red sideband case:
\begin{eqnarray}
|e_n^\pm\rangle & = &  \frac{1}{\sqrt{2}}\left
(|g,n+2\rangle\pm|e,n\rangle\right )\\
e_n^\pm & = & \pm\eta\Omega_0\sqrt{n(n+1)} \,\, .
\end{eqnarray}

The general solution to Eq. (\ref{mastereq}) may be written explicitly as
\begin{equation}
\rho(t)=e^{-iGt-\frac{\Gamma t}{2}G^2}\left (e^{{\cal J}t}\rho(0)\right
)e^{iGt-\frac{\Gamma
t}{2}G^2} \,\, , 
\end{equation}
where the superoperator in the middle of this expression is defined by the power
series expansion
for the exponential with
\begin{equation}
{\cal J}^m\rho=\Gamma^mG^m\rho G^m \,\, . 
\end{equation}
In the energy eigenstate basis we can solve Eq.(\ref{energyme}) explicitly to
give
\begin{equation}
\langle e_n^\pm|\rho(t)|e_m^\pm\rangle=\exp\left
[-\frac{i}{\hbar}t(e_n^\pm-e_m^\pm)-\frac{\Gamma
t}{2\hbar^2}(e_n^\pm-e_m^\pm)^2\right ]\langle e_n^\pm|\rho(0)
|e_m^\pm\rangle  \,\, .
\end{equation}
However it is of rather more use to exhibit the solution 
explicitly for particular initial conditions of relevance to 
the experiments. With this in mind we will assume that the 
initial state is prepared to be a particular vibrational 
energy eigenstate with the ion prepared in the 
electronic ground state,
\begin{equation}
|\psi(0)\rangle= |g,n\rangle \,\, . 
\end{equation}

\begin{figure}[htb]
\centerline{\psfig{figure=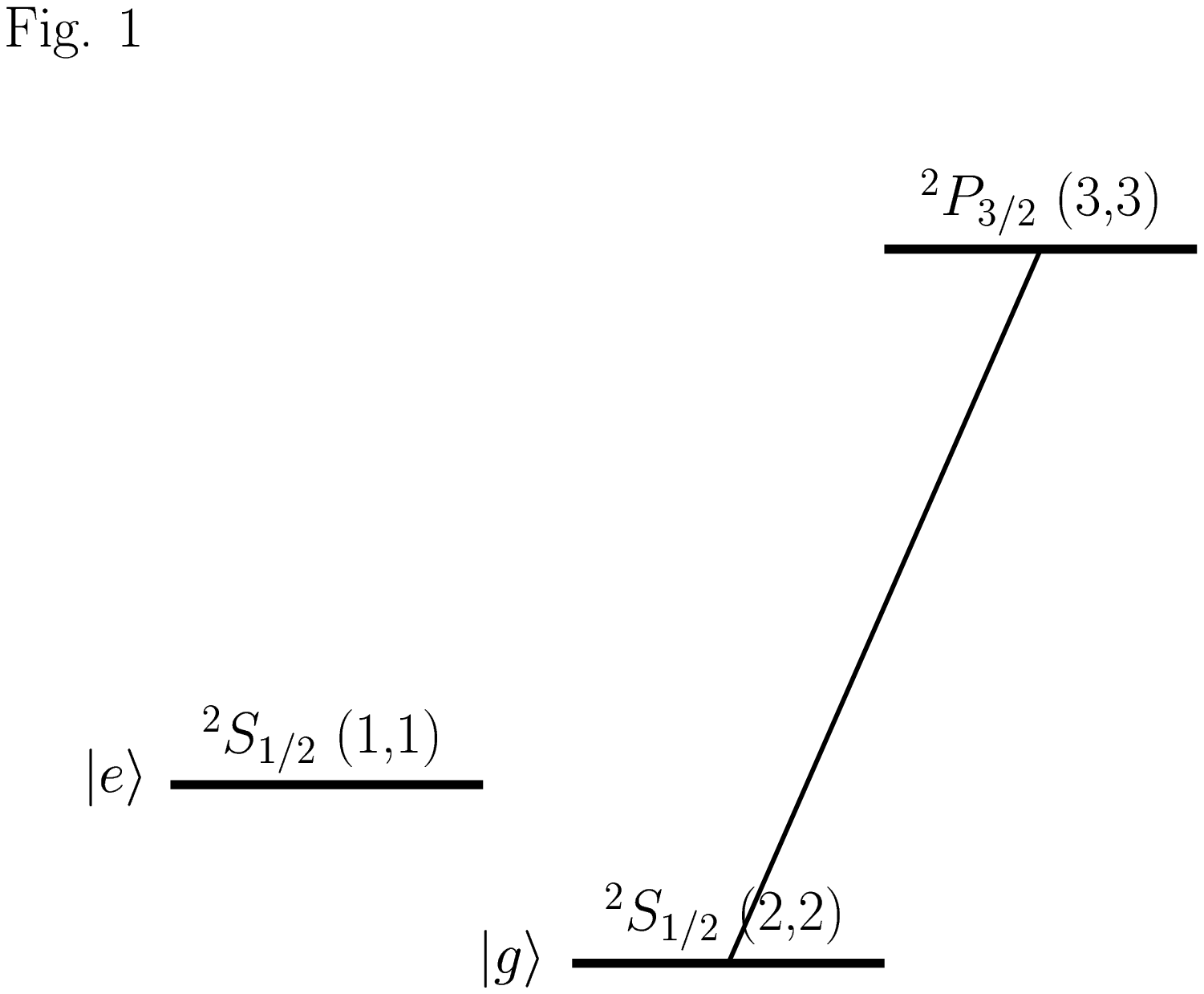,width=120mm}}
\caption{Internal level scheme of $^9$Be$^+$. The ground state $|g\rangle$
is the $2s\ ^2S_{1/2}$ ($F=2, M_F = 2$) state and the excited state $|e
\rangle$ is the $2s\ ^2S_{1/2}$ ($F=1, M_F = 1$) state. The state $|g\rangle$
is detected by applying a nearly resonant $\sigma^+$ polarized laser 
probe field to drive a strong transition between $|g\rangle$ and another 
state $^2P_{3/2}$ ($F=3, M_F = 3$) as indicated. As this other state 
can only decay back to $|g\rangle$, any fluorescence 
observed on this transition is evidence that the atom was in the 
state $|g\rangle$ at the start of the probe pulse. } 
\label{fig1}
\end{figure}

In the experiments done so far, it is possible to probe directly which
electronic state the ion occupies. In the work of Meekhof 
et al. \cite{Wineland1996} the internal state $|g\rangle$ is the 
$2s\ ^2S_{1/2}$ $(F=2,M_F=2$) state of $\ ^9$Be$^+$, and the state 
$|e\rangle$ corresponds to the $2s\ ^2S_{1/2}$ ($F=1,M_F=1$) state 
as shown in figure \ref{fig1}. The state $|g\rangle$ is
detected by applying a nearly resonant $\sigma^+$ polarized 
laser probe field to drive a strong transition between  the state
$|g\rangle$ and another state, $\ ^2P_{3/2}$ ($F=3,M_F=3$). As this 
other state can only decay back to $|g\rangle$, any fluorescence 
observed on this transition is evidence that the atom was in
the state $|g\rangle$ at the start of this probe pulse. As the 
intensity of the fluorescence on this transition is strong, it is 
almost certain to detect a photon and thus the quantum efficiency
of this state determination is near unity. In other words, this measurement
scheme realizes an almost perfect projection valued measurement 
onto the electronic state $|g\rangle$. A sequence
of probe pulses delayed a time $\tau$ after the initial state 
preparation can thus be used to build up the probability 
$P_g(t)$ to find the ion in the ground state. Of course
to do this many repeats of the experiment must be performed and 
it is not possible to track laser fluctuations
exactly one each run. The final result for $P_g(t)$ must then represent an
ensemble average over these fluctuations, and the ion dynamics 
is then described by Eq.(\ref{mastereq}).
Given $P_g(t)$, one easily sees that $P_e(t)=1-P_g(t)$. 

\begin{figure}
\centerline{\psfig{figure=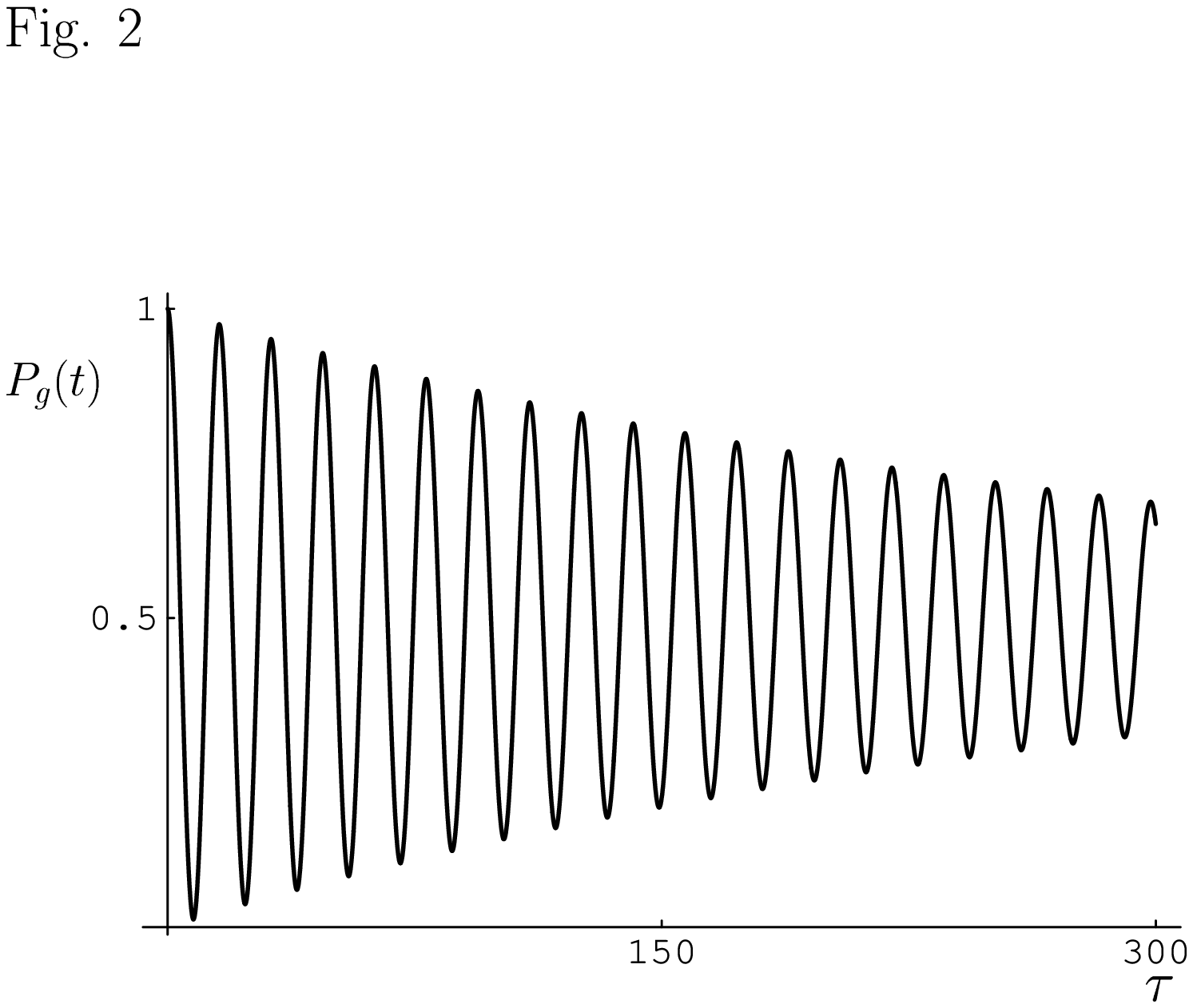,width=120mm}}
\caption{$P_g(t)$ for an initial $|g, n=0\rangle$ state of the atom driven 
by the first blue sideband, Eq.~(32). The time parameter
$\tau = \Omega_0 t$ is scaled 
with the effective Rabi frequency $\Omega_0 = 470$ kHz. The value 
for the scaled damping coefficient $\Gamma^\prime = \Gamma \Omega_0$ is 
0.041 and the one for $\eta$ is 0.2.
All the values are rough estimates from the experimental values 
given in [3].}
\label{fig2}
\end{figure}

With $\psi(0)=|g,n\rangle$, the solution for $P_g(t)$ in each 
case is as follows.
For carrier excitation:
\begin{equation}
P_g(t)=\frac{1}{2}\left [1+\exp\left\{-\frac{\Gamma
t\Omega_0^2}{2}(1+2n\eta^2)\right \}\cos\left
(\Omega_0t(1+n\eta^2)\right )\right ] \,\, . 
\end{equation}
For red side-band excitation:
\begin{equation}
P_g(t)=\frac{1}{2}\left [1+\exp\left (-2\Gamma\eta^2\Omega_0^2 nt \right
)\cos\left(2\eta\Omega_0
\sqrt{n}t \right)\right ] \,\, . 
\end{equation}
For blue sideband excitation;
\begin{equation}
P_g(t)=\frac{1}{2}\left [1+\exp\left (-2\Gamma\eta^2\Omega_0^2 (n+1)t \right
)\cos\left(2\eta\Omega_0
\sqrt{n+1}t \right)\right ] \,\, . 
\label{bluesideband}
\end{equation}
For second red sideband excitation:
\begin{equation}
P_g(t)=\frac{1}{2}\left [1+\exp\left (-2\Gamma\eta^2\Omega_0^2 n(n-1)
t \right)\cos\left(2\eta\Omega_0 \sqrt{n(n-1)}t \right)\right ] \,\, .
\end{equation}
So to test the dependence of decoherence on the excitation in the 
vibrational state experimentally, the second red sideband (or any 
higher order sideband) is a better choice than just the first order 
sidebands, since the dependence of the damping on $n$ is quadratic 
(or of even higher order for higher sidebands).
In figure \ref{fig2} we illustrate a typical result using 
Eq.~(\ref{bluesideband}) and the parameters given in \cite{Wineland1996}. 

\subsection{Phase fluctuations}

Now we investigate phase fluctuation in the laser instead of 
intensity fluctuations, i.e. we set $\Omega(t)$ in 
Eqs.~(\ref{randomham1})--(\ref{randomham}) to $\Omega_0$ and 
introduce white phase noise $\phi(t) =  \sqrt\lambda  W(t)$
instead. The means that the Hamiltonian is a nonlinear function 
of the noise which will lead to technical difficulties in 
deriving the corresponding Ito differential equation for the system
state. However for our purposes a simple transformation can simplify 
matters considerably.  We
follow \cite{dyrting1996} and use a random canonical transformation 
to get rid of the nonlinearity in the noise source. In effect this 
is an instantaneous rotation of the system through a fluctuating
angle, analogous to the standard method of transformation to an interaction
picture.  
\begin{equation}
\hat{U} = \exp \left( i \phi(t) \hat{\sigma}_+ \hat{\sigma}_- \right) \,\, . 
\end{equation}
   Thus
\begin{eqnarray}
\rho & \longrightarrow & \tilde{\rho}= \exp\left(-i \phi(t) \hat{\sigma}_+
\hat{\sigma}_- \right)  \rho \exp \left( i \phi(t) \hat{\sigma}_+ 
\hat{\sigma}_- \right) \\
H & \longrightarrow & H_0 = G -\phi(t) \hat{\sigma}_+ 
\hat{\sigma}_-  \,\,, 
\end{eqnarray}
where $G$ is one of the operators defined by Eqs.~(\ref{ggg1})--(\ref{ggg}), 
depending on which sideband the laser is tuned to. 
The reason we can use this stochastic transformation is that we are only
interested in the population of the electronic levels. As 
the generator of the unitary transformation commutes with
the population operator, the populations in the instantaneous 
transformed frame are the same as those in the original frame. 
However other moments will not be the same, and we could not easily
use the transformed state, $\tilde{\rho}$, after averaging, 
to reconstruct moments in the original frame. This transformation  
can be described covering all four cases of $G$ in one, since it only
affects  the internal state and the operators describing that 
($\hat{\sigma}_+$ and $\hat{\sigma}_-$ appear in the same order 
and form in all the interactions we consider here). 
The corresponding master equation after averaging out the noise reads
\begin{equation}
\frac{d \tilde{\rho}} {dt}  =  -i \left[ G, \tilde{\rho} \right] 
- \lambda \left[ \hat{\sigma}_+ \hat{\sigma}_-, 
\left[ \hat{\sigma}_+ \hat{\sigma}_-, 
\tilde{\rho} \right] \right] \,\, . 
\end{equation}
For the initial condition $\hat{\rho} = |g\rangle \langle g|\otimes 
|n\rangle\langle n|$ we can solve this for all sidebands. 
However, here we restrict ourselves to the red sideband since that 
gives an idea on how the effects arising from phase fluctuations
are different from intensity fluctuations. 
The probability for the atom to be in the upper state in this 
case is given by
\begin{eqnarray}
\tilde{P}_{g}(t) & = & \frac{1}{2} \left[1 + \cos(\tilde{\omega}_n) 
\exp(-(\lambda/2) t )
+ \frac{\lambda}{2\tilde{\omega}_n} 
 \sin (\tilde{\omega}_n)  
 \exp(-(\lambda/2) t ) \right] \,\, (\mbox{red sideband}), 
\end{eqnarray}
with
\begin{equation}
\tilde{\omega}_n = \sqrt{4 \eta^2 \Omega_0^2 n - \lambda^2/4} \,\, . 
\end{equation}
So here the effective Rabi frequency depends on the 
coherence decay rate and not vice 
versa as in the case of intensity fluctuations.

\section{Discussion}

We have shown how to model fluctuations in the laser causing the 
interaction between center-of-mass motion and internal states 
in ion traps. Intensity fluctuations lead to decoherence processes 
which depend on the kind of interaction the laser is causing. 

What is the value for $\Gamma$ for fluctuations in 
the laser intensity? A rough estimate from the figures given in
\cite{Wineland1996},
using the quoted experimental values, leads to $\Gamma \approx 
1.4 \cdot 10^{-8} s$. But this is really a very rough estimation. 
If we define the fractional error by the quotient of the r.m.s. and the
pulse area, 
\begin{equation}
\frac{\mbox{r.m.s.}}{A(T)}  = \frac{\Omega_0 \sqrt{\Gamma T}}{\Omega_0 T}
= \sqrt{\frac{\Gamma}{T}}\,\, ,
\end{equation}
a fractional error of one percent leads to $\Gamma \approx 10^{-10}$. 
With a fractional error of ten percent, however, we get $\Gamma \approx
10^{-8}$ and we are in the range of the roughly estimated value for
$\Gamma$. 

In \cite{Wineland1996} they experimentally estimate the $n$-dependence 
of their damping which they fit with 
\begin{equation}
P_g(t) = \frac{1}{2} \left[ 1 + \sum_{n=0} P_n\cos(2\Omega_{n, n+1}t)
e^{-\gamma_n t} \right]
\end{equation}
to be of the form 
\begin{equation}
\gamma_n = \gamma_o (n+1)^{0.7} \,\,. 
\end{equation}
We derive the coefficient to be 0.5 instead of 0.7 if the decoherence 
is just due to intensity fluctuations alone. However there are other 
sources of decoherence due, for example, to fluctuations in the 
trap potential itself. These fluctuations lead to, among other
things, a fluctuating trap center point and will be addressed in a future
publication.  

\section*{Acknowledgments}
S.~S. gratefully acknowledges financial
support from a ``DAAD Doktorandenstipendium
im Rahmen des gemeinsamen Hochschulsonderprogramms III von Bund und
L\"andern'' and from the Center of Laser Science. 
%
%

\newpage

\thebibliography{99}
\bibitem{Wineland1997}D.J.Wineland, C. Monroe, W.M. Itano, D. Liebfried, 
B.King, and D.M. Meekhof,
"Experimental issues in coherent quantum-state manipulation of trapped atomic
ions", submitted to
Rev. Mod. Phys. (1997). 

\bibitem{Monroe1996}C. Monroe, D.M.  Meekof, B.E. King, and D.J. Wineland, 
Science, {\bf 272}, 1131 (1996).

\bibitem{Wineland1996}D.M. Meekof, C. Monroe, B.E. King, W.M. Itano, and 
D. J. Wineland, Phys. Rev. Lett. {\bf 76}, 1796 (1996) and Phys. Rev. Lett. 
{\bf 77}, 2346 (Erratum) (1996).

\bibitem{Cirac1995} J.~I.~Cirac and P.~Zoller, Phys. Rev. Lett. {\bf 74}, 
4094 (1995).

\bibitem{Monroe1995} C. Monroe, D.M. Meekhof, B.E. King, W.M. Itano, 
and D.J. Wineland, Phys. Rev. Lett. {\bf 75}, 4714 (1995).

\bibitem{Gardiner1985}C.W. Gardiner,{\it Handbook of Stochastic Processes for
Physics, Chemistry
and the Natural Sciences}, (Springer-Verlag, Berlin, 1985).

\bibitem{JC} B.W. Shore and P.L. Knight, J. Mod. Opt. {\bf 40}, 1195 (1993).

\bibitem{dyrting1996}S. Dyrting and G.J. Milburn, Quantum Semiclass. Opt.
{\bf 8}, 541 (1996).

\bibitem{milburn1991}G.J. Milburn, Phys. Rev. A {\bf 44}, 5401 (1991).

\bibitem{Moya}H. Moya-Cessa, V. Buzek, M.S. Kim, and P.L. Knight, Phys. Rev.
A {\bf 48}, 3900 (1993).  

\bibitem{Kuang} L.-M. Kuang, X. Chen, G.-H. Chen, and M.-L. Ge, 
Phys. Rev. A {\bf 56}, 3139 (1997).
\end{document}